\documentclass[12pt]{article}
\usepackage{graphicx}
\usepackage{cite}
\newcommand{\bm}[1]{\mbox{\boldmath $#1$}}
\newcommand{\fnd}[2]{\frac{\textstyle #1}{\textstyle #2}}
\newcommand{\icgp}[1]{\resizebox{6.0cm}{!}{\includegraphics{#1}}}
\begin{document}
\title{Multichannel calculation of $D_{s}^{\ast}$ vector states
and the $D^+_{sJ}(2632)$ resonance.}
\author{
Eef van Beveren\\
{\normalsize\it Centro de F\'{\i}sica Te\'{o}rica}\\
{\normalsize\it Departamento de F\'{\i}sica, Universidade de Coimbra}\\
{\normalsize\it P-3000 Coimbra, Portugal}\\
{\small http://cft.fis.uc.pt/eef}\\ [.3cm]
\and
George Rupp\\
{\normalsize\it Centro de F\'{\i}sica das Interac\c{c}\~{o}es Fundamentais}\\
{\normalsize\it Instituto Superior T\'{e}cnico, Edif\'{\i}cio Ci\^{e}ncia}\\
{\normalsize\it P-1049-001 Lisboa Codex, Portugal}\\
{\small george@ajax.ist.utl.pt}\\ [.3cm]
{\small PACS number(s): 14.40.Lb, 13.25.Ft, 12.39.Pn, 13.75.Lb}\\ [.3cm]
{\small hep-ph/0407281}
}

\maketitle

\begin{abstract}
We study bound states below threshold and resonances above threshold
in the $D^{0}K^{+}$ and $D_{s}^{+}\eta$ systems,
using a many-coupled-channel model for non-exotic meson-meson scattering
applied to states with the quantum numbers of
$c\bar{s}$ quark-antiquark vector mesons.
We fit the ground state at 2.112 GeV, whence the lowest resonances in
$D^{0}K^{+}$ come out at 2.61, 2.72, 3.03, and 3.08 GeV.
The resonance at 2.61 GeV acquires a width of about 8 MeV, while its partial
P-wave cross section is up to six times larger in $D_{s}\eta$ than in
$D^{0}K^{+}$, provided a mechanism accounting for
Okubo-Zweig-Iizuka--forbidden decays is included.
The latter finding is in agreement with the observations of
the SELEX collaboration \cite{HEPEX0406045} with respect to the recently
reported $D^+_{sJ}(2632)$ resonance.
Therefore, we conclude that the $D^+_{sJ}(2632)$
is probably the first recurrence of the $D_{s}^{\ast}(2112)$ meson.
\end{abstract}
\clearpage

\section{Introduction}

The very fortunate discovery \cite{HEPEX0406045} of a new resonance
in both the $D^{0}K^{+}$ and $D_{s}^{+}\eta$ channels
has been awaited for over two decades \cite{DASP}.
Excitations of $J^{P}=1^{-}$ $c\bar{s}$ states have been foreseen in the
past, but with higher masses. For instance,
in Ref.~\cite{PRD32p189} two states were forecast
at 2.73 and 2.90 GeV, one for each of the possible angular configurations
$S$ and $D$, whereas in Ref.~\cite{PRD36p1465} these states were predicted
at 2.773 and 2.813 GeV.
Nowadays, a mass of 2.632 GeV is of no surprise.
Just consider the mass differences between the ground states and first
radial excitations of other $J^{P}=1^{-}$ mesons,
which vary from 0.52 GeV for $K^{\ast}$
to 0.59 GeV for $c\bar{c}$.
The $D^+_{sJ}(2632)$ resonance, being 0.52 GeV heavier than
the $D_{s}^{\ast}$ ground state,
could thus perfectly well turn out be the first radial excitation of the
$D_{s}^{\ast}(2112)$ meson.
However, the branching ratio of its two decay modes, i.e.,
$D^{0}K^{+}$ to $D_{s}^{+}\eta$,
with the latter one dominating by a factor of six,
seems to be problematic \cite{HEPEX0406045,HEPPH0407120}.
Also its decay width (smaller than 17 MeV \cite{HEPEX0406045})
is smaller than expected for a $c\bar{s}$ resonance with sufficient phase
space in a two-meson channel observing the Okubo-Zweig-Iizuka (OZI) \cite{OZI}
rule.

This small decay width, and the puzzling branching ratio to
$D^{0}K^{+}$ and $D_{s}^{+}\eta$, are explained by Maiani {\em et al.}
\/\cite{HEPPH0407025} assuming a dominantly
$\left[ cd\right]\left[\bar{d}\bar{s}\right]$ four-quark configuration,
whereas Liu {\em et al.} \/\cite{HEPPH0407157} argue that
the $D^+_{sJ}(2632)$ resonance may be a member of a flavor 15-plet,
which is a mixture of $\left[ cu\right]\left[\bar{u}\bar{s}\right]$,
$\left[ cd\right]\left[\bar{d}\bar{s}\right]$, and
$\left[ cs\right]\left[\bar{s}\bar{s}\right]$.
Other tetraquark proposals can be found in Refs.~\cite{HEPPH0407062}
and \cite{HEPPH0407088}, namely
$\left[ c\bar{s}\right]\left[u\bar{u}\!+\!d\bar{d}\right]$ and
$\left[ cd\right]\left[\bar{d}\bar{s}\right]$, respectively.
A radial excitation of the $D_{s}^{\ast}$ is suggested
in Refs.~\cite{HEPPH0407091,HEPPH0407120}, among other possibilities.

Here, we shall show that the observed properties are compatible
with the predictions of a multichannel quark-meson model, designed to describe
mesonic resonances in terms of systems of coupled $q\bar{q}$ and two-meson
states \cite{HEPPH0211411}.
However, the study of non-ground-state $D_{s}^\ast$ vector mesons
requires a more elaborate approach than the one previously applied by us to
scalar mesons \cite{EPJC22p493}, as we shall explain in the following.

In the first place, $J^{P}\!=\!1^{-}$ systems
consisting of a charm quark and a strange antiquark
may appear in two different orbital excitations,
namely $\ell =0$ and $\ell =2$.
This implies that instead of one $c\bar{s}$ channel,
as in the case of the $D_{s0}^{\ast}(2317)$ resonance \cite{PRL91p012003},
we must now consider two such channels.

Secondly, higher radial and angular excitations of $c\bar{s}$
can decay into a variety of different OZI-allowed two-meson channels,
while the influence of nearby closed two-meson channels must be taken into
account, too.
In Table~\ref{MMchannels}, we summarize
the twenty-eight mesonic decay channels of the $c\bar{s}$ system
that we consider of importance in the energy region of interest.

Thirdly, the phenomenological Resonance-Spectrum-Expansion (RSE) formalism
employed in Refs.~\cite{EPJC22p493,PRL91p012003}
is not practicable in the present case, exactly
because of the proliferation of relevant two-meson channels involved.
In the RSE method, one term in the expansion accounts for one radial or
angular $q\bar{q}$ channel coupled to one meson-meson channel.
Now, each term comes with a different parameter for each combination of
one of the $c\bar{s}$ channels and one of the two-meson channels.
Thus imagining four terms in the expansion,
then with two channels in the $c\bar{s}$ sector
and twenty-eight channels in the two-meson sector
we would end up with many parameters:
clearly enough to fit anything, but not very useful for predictions.

Nonetheless, with a definite choice for the confinement potential,
these ambiguities are readily removed.
Here, we opt for a harmonic oscillator, which has the additional advantage
that the relative decay couplings, given in Table~\ref{MMchannels},
can be determined through the formalism developed in
Ref.~\cite{recouple}.

Since, moreover, we consider only one flavor state
with reasonably heavy decay products,
it is justified to work in the spherical-delta-shell approximation
for the $^3\!P_0$ communication between the $c\bar{s}$ and two-meson sectors.
Thus, we shall employ here the model developed in Ref.~\cite{PRD21p772} for
$c\bar{c}$ and $b\bar{b}$ states, and generalized to other flavors in
Ref.~\cite{PRD27p1527}.
The relative couplings are determined with the formalism
from Ref.~\cite{recouple},
where it is assumed that each decay product takes over one
of the original constituent quarks.
Other processes are here classified as OZI forbidden.
\begin{table}[htbp]
\begin{center}
\begin{tabular}{||c|c|c|cc||}
\hline\hline & & & & \\ [-10pt]
channel & threshold & &
\multicolumn{2}{c||}{relative couplings}\\ [3pt]
& GeV & $(L,S)$ & to $\ell =0$ & to $\ell =2$\\
\hline & & & & \\ [-10pt]
$D^{0}$ - $K^{+}$ & 2.358 & (1,0) & 1/72 & 1/216\\
\hline & & & & \\ [-12pt]
$D^{+}$ - $K^{0}$ & 2.367 & (1,0) & 1/72 & 1/216\\
\hline & & & & \\ [-12pt]
$D^{\ast 0}$ - $K^{+}$ & 2.500 & (1,1) & 1/36 & 1/432\\
\hline & & & & \\ [-12pt]
$D^{\ast +}$ - $K^{0}$ & 2.508 & (1,1) &  1/36 & 1/432\\
\hline & & & & \\ [-12pt]
$D_{s}^{+}$ - $\eta$ & 2.516 & (1,0) & 1/108 & 1/288\\
\hline & & & & \\ [-12pt]
$D^{\ast +}_{s}$ - $\eta$ & 2.660 & (1,1) &  1/54 & 1/648\\
\hline & & & & \\ [-12pt]
$D^{0}$ - $K^{\ast +}$ & 2.756 & (1,1) &  1/36 & 1/432\\
\hline & & & & \\ [-12pt]
$D^{+}$ - $K^{\ast 0}$ & 2.766 & (1,1) &  1/36 & 1/432\\
\hline & & & & \\ [-12pt]
$K^{\ast}_{0}(800)$ - $D^{\ast}$ & 2.838 & (0,1) & 1/24 & 1/216\\
& & (2,1) &  & 1/216\\
\hline & & & & \\ [-12pt]
$D^{\ast 0}$ - $K^{\ast +}$ & 2.898 & (1,0) & 1/216 & 1/648\\
& & (1,2) & 5/54 & 1/3240\\
& & (3,2) &  & 7/120\\
\hline & & & & \\ [-12pt]
$D^{\ast +}$ - $K^{\ast 0}$ & 2.906 & (1,0) & 1/216 & 1/648\\
& & (1,2) & 5/54 & 1/3240\\
& & (3,2) &  & 7/120\\
\hline & & & & \\ [-12pt]
$D_{s}^{+}$ - $\eta '$ & 2.926 & (1,0) & 1/216 & 1/648\\
\hline & & & & \\ [-12pt]
$D_{s}$ - $\phi$ & 2.988 & (1,1) & 1/36 & 1/432\\
\hline & & & & \\ [-12pt]
$D^{\ast +}_{s}$ - $\eta '$ & 3.070 & (1,1) & 1/108 & 1/1296\\
\hline & & & & \\ [-12pt]
$f_{0}(980)$ - $D^{\ast}_{s}$ & 3.092 & (0,1) & 1/48 & 1/432\\
& & (2,1) &  & 1/432\\
\hline & & & & \\ [-12pt]
$D^{\ast}_{s}$ - $\phi$ & 3.132 & (1,0) & 1/216 & 1/648\\
& & (1,2) & 5/54 & 1/3240\\
& & (3,2) &  & 7/120\\
\hline & & & & \\ [-12pt]
$D_{0}^*(2308)$ - $K^{\ast}$ & 3.201 & (0,1) & 1/24 & 1/216\\
& & (2,1) &  & 1/216\\
\hline & & & & \\ [-12pt]
$D^{\ast}_{s0}(2317)$ - $\phi$ & 3.337 & (0,1) & 1/48 & 1/432\\
& & (2,1) &  & 1/432\\
\hline\hline
\end{tabular}
\end{center}
\caption{The pseudoscalar-pseudoscalar, pseudoscalar-vector,
vector-vector, and scalar-vector real and virtual
OZI-allowed two-meson channels, for $L\!=0\!$, 1, 2, 3,
considered throughout this work, their thresholds, and
relative couplings squared to each of the two $c\bar{s}$ channels.
For $\eta$ and $\eta '$ we assume flavor octet and singlet, respectively.
Experimental masses are taken from Ref.~\cite{PLB592p1}, except for the
$D_0^*$(2308) \cite{PRD69p112002} (see also Ref.~\cite{PRL91p012003}),
and the $K_0^*$(800) (or $\kappa$) \cite{PLB592p1}. For the latter resonance,
we choose the peak mass of 830 MeV from Ref.~\cite{EPJC22p493}.}
\label{MMchannels}
\end{table}
\clearpage

\section{Inelastic meson-meson scattering}

The $28\times 28$ scattering matrix $S$,
as a function of the total center-of-mass energy $\sqrt{s}$,
has the familiar form
\begin{equation}
S\left(\sqrt{s}\right)\; =\;
k^{1/2}\;\left[ 1-iK\right]^{-1}\;\left[ 1+iK\right]\; k^{-1/2}
\;\;\; ,
\label{Smatrix}
\end{equation}
where $k$ and $L$ can be represented by diagonal matrices containing
the linear and angular momenta of each of the two-meson channels, respectively.
The elements of the matrix $k$ are determined through the kinematically
relativistic expression
\begin{equation}
4s k_{i}^{2}\left(\sqrt{s}\right)\! =\!
\left[ s-\left( m_{1i}+m_{2i}\right)^{2}\right]\!
\left[ s-\left( m_{1i}-m_{2i}\right)^{2}\right]
,
\label{kmatrix}
\end{equation}
where $m_{1i}$ and $m_{2i}$ stand for the meson masses involved in
the $i$-th two-meson channel.
The $28\times 28$ inverse cotangent matrix $K$,
from which scattering phase shifts can be obtained,
is defined by
\begin{equation}
K\left(\sqrt{s}\right)\; =\;
-\left[ 1-k^{-1}M\; JXN\right]^{-1}\;\left[ k^{-1}M\; JXJ\right]
\;\;\; .
\label{Kmatrix}
\end{equation}
In this equation, the diagonal matrices $J$ and $N$ contain
the two linearly independent Bessel and Neumann scattering solutions
above threshold, respectively, or their analytic continuations below threshold,
for each of the two-meson channels.
The diagonal elements of the reduced-mass matrix $M$
are calculated by using the relativistic formula
\begin{equation}
4s^{3/2} M_{i}\left(\sqrt{s}\right)\! =\!
\left[ s^{2}-
\left( m_{1i}+m_{2i}\right)^{2}
\left( m_{1i}-m_{2i}\right)^{2}
\right]
.
\label{Mmatrix}
\end{equation}
The $28\times 28$ matrix $X$ introduced in Eq.~(\ref{Kmatrix}) is defined as
\begin{equation}
X\left(\sqrt{s}\right)\; =\;
4\lambda^{2}\; V^{T}AV
\;\;\; ,
\label{Xmatrix}
\end{equation}
where $\lambda$ is the overall coupling constant related to
quark-pair creation, and the $2\times 28$ matrix $V$
contains the relative coupling constants given in
Table~\ref{MMchannels}.
The diagonal $2\times 2$ matrix $A$ contains products of
the two linearly independent solutions of the harmonic oscillator.
It is this matrix for which an RSE expansion can be formulated.

Except for the matrix $V$, all matrices are diagonal, because we do not
consider direct interactions in either of the two sectors.
More details can be found in Ref.~\cite{PRD27p1527}.

From Ref.~\cite{PRD27p1527}, we adopt here
the effective constituent quark masses
$m_{c}\!=\!1.562$ GeV and $m_{s}\!=\!0.508$ GeV,
and the oscillator frequency $\omega\!=\!0.19$ GeV.
However, we cannot use the scaled dimensionless radius $\rho_{0}$
from Ref.~\cite{PRD27p1527}, since it corresponds to the maxima of different
transition potentials \cite{recouple}, not their effective radii.
In Ref.~\cite{EPJC22p493} we used 5 GeV$^{-1}$ for $K\pi$ in  a $P$ wave,
which gives $\rho_{0}\!\approx\!1$,
and 3.2 GeV$^{-1}$ for $K\pi$ in an $S$ wave,
yielding $\rho_{0}\!=\!0.66$.
For simplicity, we choose here $\rho_{0}\!=\!0.8$, as a kind of average value
for the different waves under consideration (see Table~\ref{MMchannels}),
corresponding to a distance of 0.58 fm.
The overall coupling $\lambda$ we vary such that
the ground state of the spectrum, i.e., the $D_{s}^{\ast}$(2112),
gets its experimental \cite{PLB592p1} mass.
\clearpage

\section{OZI-forbidden decays}

Our initial findings, with OZI-allowed channels only, are the following.
We obtain a narrow resonance at about 2.6 GeV.
Its width and position depend on the invariant radius $\rho_{0}$.
For the above value of $\rho_{0}\!=\!0.8$, the resonance comes out
at 2.613 GeV, and has a width of  6.5 MeV. Hence, contrary to what might
be expected from naive perturbative calculations, we obtain a {\em narrow}
\/resonance for a system
having more than enough phase space for OZI-allowed decay.
The reason appears to be twofold. First of all, most of the probability of
$^{3}\!P_{0}$ pair creation goes into closed channels, which only give rise to
a real mass shift. Secondly, the first radial excitation we are assuming for
the $D^+_{sJ}(2632)$, albeit mixed with the $1^3\!D_1$ state, has one node
(see also Ref.~\cite{HEPPH0407091}), which cannot lie far away from the the
delta-shell radius $\rho_0$, on the basis of general size arguments. In the
more realistic picture of Ref.~\cite{PRD27p1527}, where a smooth transition
potential was used, the convolution of this potential with the sign-changing
wave function of the ``bare'' $D^+_{sJ}(2632)$  will lead to a partial
cancellation. So the total width due to OZI-allowed decays could be
considerably smaller than what is generally assumed.

Now, once we have a full solution of the coupled-channel problem,
i.e., the complete $S$ matrix,
we may determine branching ratios for other types of decay modes,
namely OZI-forbidden processes.
However, one of these happens to contribute to the very $D_{s}^{+}\eta$ mode,
already included in the list of OZI-allowed decay channels. Notice that an
OZI decay to $D_{s}^{+}\eta$ can only proceed via $s\bar{s}$ pair creation,
whereas the non-OZI decay may take place through both $u\bar{u}$/$d\bar{d}$ and
$s\bar{s}$ creation.
Precise values for such OZI-forbidden decays have to be estimated, since we do
not dispose of a lot of experimental data on such processes when
alternative OZI-allowed modes exist, too.
Here, we estimate that up to 4--5 percent of the
total probability of strong decay contributes to OZI-forbidden decays,
which we account for in an effective way
by correspondingly increasing the coupling constant of
the OZI-allowed $D_{s}^{+}\eta$ channel.
Note that this estimate of 4--5\% seems quite reasonable
on the basis of experimentally known
\cite{PLB592p1} OZI-forbidden two-particle decays of mesons, keeping also in
mind that the mentioned suppression due to the nodal structure of the wave
function may not apply to non-OZI decays. In Table~\ref{Bratio},
we present the results, after refitting the overall coupling constant to the
$D_{s}^{\ast}$(2112) mass, for the effect of the non-OZI $D_{s}^{+}\eta$ decay
mode on the $D_{s}^{\ast'}\!\to\!D^{0}K^{+}\,/\,D_{s}^{\ast'}\!\to\!D_{s}\eta$
branching ratio, as well as the total width.
We see that the surprising branching ratio
$D_{s}^{\ast'}\!\to\!D_s\eta\,/\,D_{s}^{\ast'}\!\to\!D^0K^+\approx6$ observed
by the SELEX collaboration \cite{HEPEX0406045} is reproduced for a non-OZI
contamination of about 4.3\%, implying a total width of some 8 MeV,
which is also compatible with the data.
Of course, the latter experimental findings need further confirmation.
What our results show is that such a scenario is plausible.
\begin{table}[htbp]
\begin{center}
\begin{tabular}{||c|c|c|c|c||}
\hline\hline & & & & \\ [-10pt]
non-OZI ($\%$) & peak (GeV) &
$\fnd{D_ {s}^{\ast '}\to D^{0}K^{+}}{D_ {s}^{\ast '}\to D_{s}\eta}$ &
$\Gamma$ (MeV) & $\lambda$\\
& & & &\\ [-10pt]
\hline & & & &\\ [-10pt]
0 & 2.613 & 10.46 & 6.5 & 2.67\\
0.31 & 2.613 & 4.47 & 6.6 & 2.66\\
0.62 & 2.613 & 2.52 & 6.6 & 2.65\\
0.93 & 2.612 & 1.69 & 6.9 & 2.63\\
1.23 & 2.612 & 1.18 & 7.0 & 2.62\\
1.85 & 2.611 & 0.67 & 7.2 & 2.60\\
2.47 & 2.611 & 0.43 & 7.4 & 2.58\\
4.32 & 2.610 & 0.17 & 7.9 & 2.51\\
\hline\hline
\end{tabular}
\end{center}
\caption{The $D_ {s}^{\ast '}$ central resonance position,
the branching ratio of $D_ {s}^{\ast '}\!\to\!D^{0}K^{+}$
to $D_{s}^{\ast '}\!\to\!D_{s}\eta$,
the total width of the $D_ {s}^{\ast '}$,
and the effective overall coupling constant,
as a function of the percentage of non-OZI
w.r.t.\ OZI hadronic decay modes.}
\label{Bratio}
\end{table}
\clearpage

\section{\bm{P}-wave cross sections for \bm{c\bar{s}} decay.}

In this section,
we present some figures for the computed partial cross sections,
corresponding to the case which reproduces the SELEX decay-rate ratio
(4.32\% in Table~\ref{Bratio}).

\subsection{The \bm{D_{s}^{\ast '}} resonance.}

\begin{figure}[htbp]
\begin{center}
\begin{tabular}{cc}
\icgp{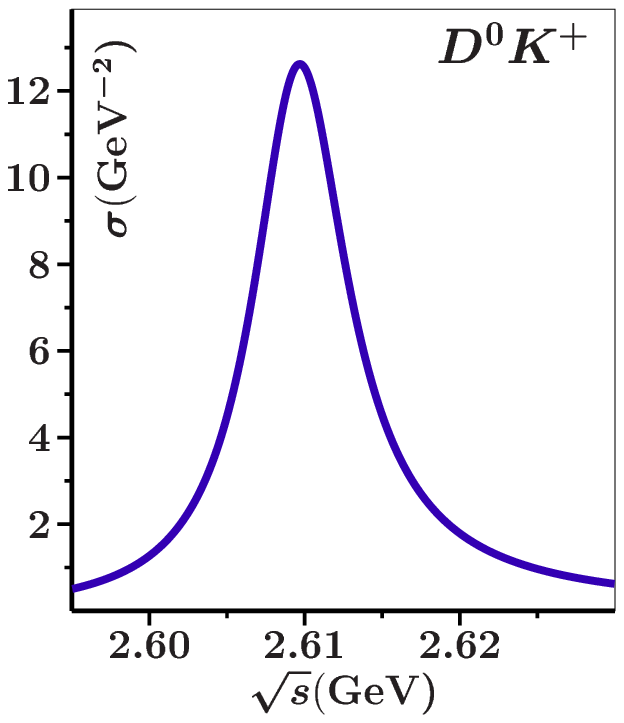} & \icgp{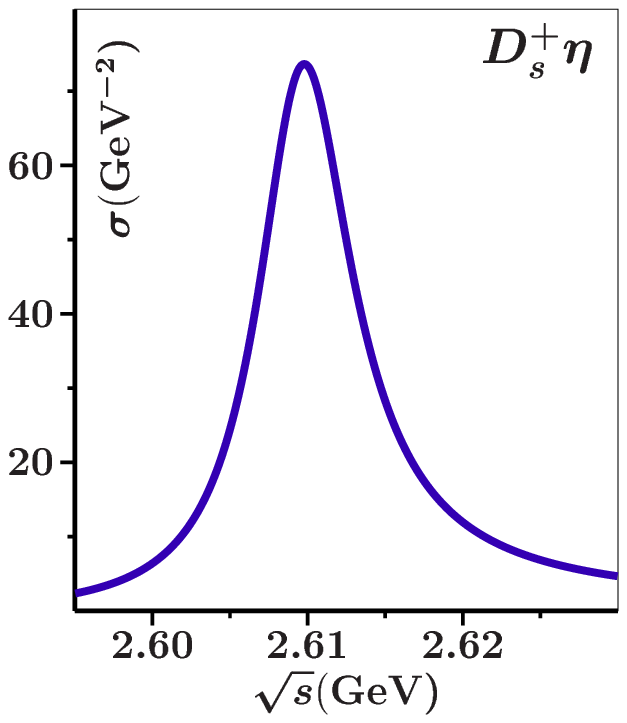}\\
(a) & (b)
\end{tabular}
\end{center}
\vspace{-15pt}
\caption[]{Theoretical cross section $\sigma$
for $D^{0}K^{+}$ (a) and $D_{s}\eta$ (b) $P$-wave scattering,
from a multi-coupled-channel calculation
(model parameters taken from Ref.~\cite{PRD27p1527},
except for the transition-potential parameters).}
\label{DKDseta}
\end{figure}

In Fig.~\ref{DKDseta}, we depict the theoretical cross
sections for $D^{0}K^{+}$ and $D_{s}^{+}\eta$ in the total-invariant-mass
($\sqrt{s}$) region of 2.6 GeV. We observe a resonance with a central mass
about 20 MeV below the SELEX value, and a width of 7.9 MeV.
Moreover, we see that about 6 times more $D_{s}^{+}\eta$ pairs are produced
than $D^{0}K^{+}$ pairs.
This resonance has here vector-meson quantum numbers, i.e., $J^{P}=1^{-}$.
\clearpage

\subsection{Further \bm{P}-wave \bm{D^{0}K^{+}} cross sections.}

\begin{figure}[htbp]
\begin{center}
\begin{tabular}{c}
\icgp{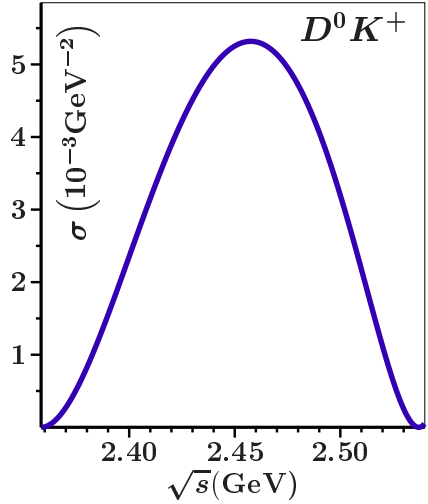}\\
\end{tabular}
\end{center}
\vspace{-15pt}
\caption[]{A broad but tiny theoretical signal in $D^{0}K^{+}$ $P$-wave
scattering below the $D_{s}\eta$ threshold.}
\label{Deffect}
\end{figure}

\begin{figure}[htbp]
\begin{center}
\begin{tabular}{cc}
\icgp{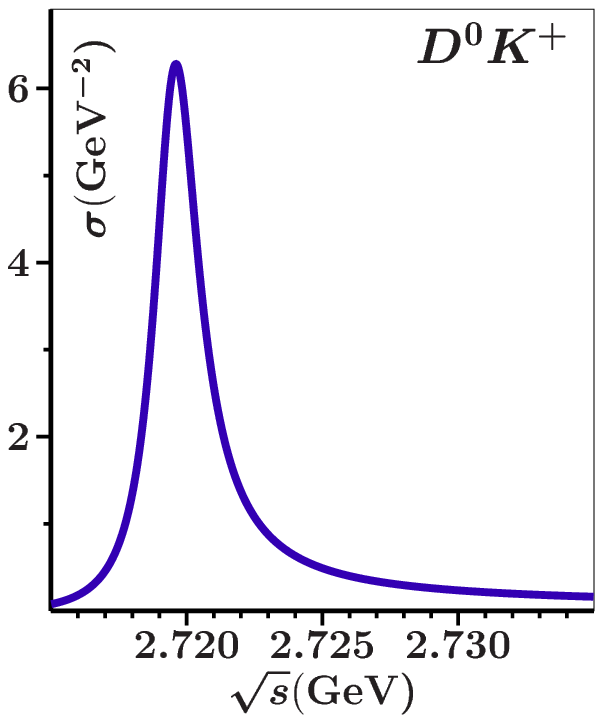} & \icgp{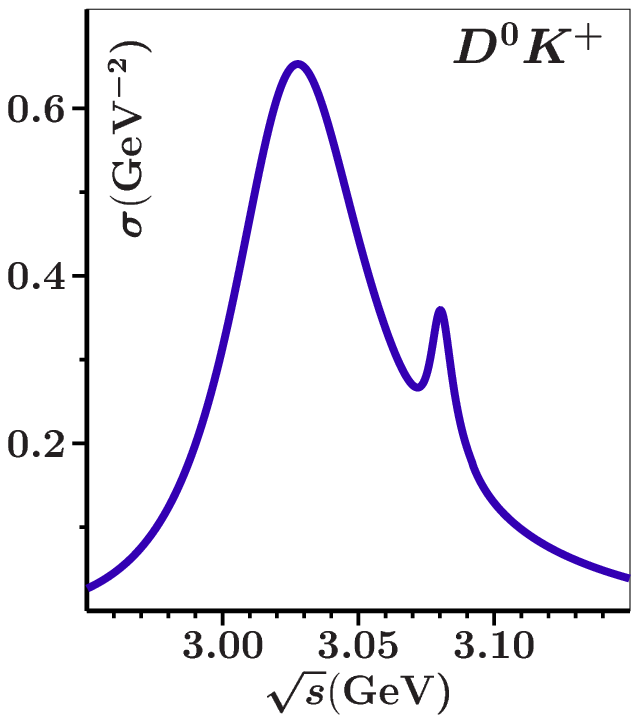}\\
(a) & (b)
\end{tabular}
\end{center}
\vspace{-15pt}
\caption[]{Details of the predicted resonances
with central masses at 2.72 GeV (a); 3.03 and 3.08 GeV (b).
Their widths are about 2, 60, and 8 MeV, respectively.}
\label{2S2D3S}
\end{figure}

In Figs.~\ref{Deffect} and \ref{2S2D3S}
we show other structures we find in $D^{0}K^{+}$ $P$-wave scattering,
in the mass region from threshold up to 3.1 GeV.

Close to the $D^{0}K^{+}$ threshold and well below
the $D_{s}^{+}\eta$ threshold,
we find a structure with a width of some 100 MeV (see Fig.~\ref{Deffect}).
Notice, however, that the peak cross section is only
0.005 GeV$^{-2}$. Whether such a weak signal can be observed in
experiment looks doubtful. This threshold effect is clearly due to the
bound-state pole of the $D_{s}^{\ast}$(2112), and does not correspond to a
resonance, since we do not encounter any additinal pole in this energy region.
\clearpage

For energies higher than the $D_{s}^{\ast '}$ mass,
we find three signals (see Fig.~\ref{2S2D3S}):
two narrow resonances at 2.72 GeV and 3.08 GeV,
and one broad signal at 3.03 GeV.
OZI-forbidden decay modes may very well broaden the two narrow structures.
So what exactly can be found in experiment, is not entirely clear from our
predictions.
We only assure that narrow companions to the $D^+_{sJ}(2632)$ should exist.
Their positions are predicted here within the model error of some 50 MeV.
The broader resonance at 3.03 GeV, with a width of 50--60 MeV,
should be reasonably easy to observe experimentally.
\clearpage

\section{Conclusions}

The experimental confirmation of at least some of the here predicted signals
will contribute strongly to our understanding of the $c\bar{s}$ spectrum.

\section*{Acknowledgments}

This work was supported in part by the
{\it Funda\c{c}\~{a}o para a Ci\^{e}ncia e a Tecnologia}
of the {\it Minist\'{e}rio da
Ci\^{e}ncia e do Ensino Superior} \/of Portugal,
under contract numbers
POCTI/\-35304/\-FIS/\-2000
and
POCTI/\-FNU/\-49555/\-2002.


\clearpage


\begin{thebibliography}{19}
\bibitem{HEPEX0406045}
A.~V.~Evdokimov  [SELEX Collaboration],
arXiv:hep-ex/0406045.

\bibitem{DASP}
R.~Brandelik {\it et al.}  [DASP Collaboration],
Phys.\ Lett.\ B {\bf 80}, 412 (1979);
{\bf 70}, 132 (1977)
(DESY reports 78/63 and 77/44, respectively).

\bibitem{PRD32p189}
Stephen Godfrey and Nathan Isgur,
Phys.\ Rev.\ D {\bf 32}, 189 (1985).

\bibitem{PRD36p1465}
D.~D.~Brayshaw,
Phys.\ Rev.\ D {\bf 36}, 1465 (1987).

\bibitem{HEPPH0407120}
T.~Barnes, F.~E.~Close, J.~J.~Dudek, S.~Godfrey, and E.~S.~Swanson,
arXiv:hep-ph/0407120.

\bibitem{OZI}
S.~Okubo,
Phys.\ Lett.\ {\bf 5}, 165 (1963);
G.~Zweig,
CERN Reports TH-401 and TH-412;
see also
{\it Developments in the Quark Theory of Hadrons}, Vol. 1, pp.\ 22--101 (1981),
edited by D.~B.~Lichtenberg and S.~P.~Rosen;
J.~Iizuka, K.~Okada, and O.~Shito,
Prog.\ Theor.\ Phys.\  {\bf 35}, 1061 (1966).

\bibitem{HEPPH0407025}
L.~Maiani, F.~Piccinini, A.~D.~Polosa, and V.~Riquer,
arXiv:hep-ph/0407025.

\bibitem{HEPPH0407157}
Y.~R.~Liu, Y.~B.~Dai, C.~Liu, and S.~L.~Zhu,
arXiv:hep-ph/0407157.

\bibitem{HEPPH0407062}
Y.~Q.~Chen and X.~Q.~Li,
arXiv:hep-ph/0407062.

\bibitem{HEPPH0407088}
B.~Nicolescu and J.~P.~B.~de Melo,
arXiv:hep-ph/0407088.

\bibitem{HEPPH0407091}
K.~T.~Chao,
arXiv:hep-ph/0407091.

\bibitem{HEPPH0211411}
E.~van~Beveren, G.~Rupp, N.~Petropoulos, and F.~Kleefeld,
AIP Conf.\ Proc.\  {\bf 660}, 353--366 (2003)
[arXiv:hep-ph/0211411].

\bibitem{EPJC22p493}
E.~van Beveren and G.~Rupp,
Eur.\ Phys.\ J.\ C {\bf 22}, 493 (2001)
[arXiv:hep-ex/0106077].

\bibitem{PRL91p012003}
E.~van Beveren and G.~Rupp,
Phys.\ Rev.\ Lett.\  {\bf 91}, 012003 (2003)
[arXiv:hep-ph/0305035].

\bibitem{recouple}
E.~van Beveren,
Z.\ Phys.\ C {\bf 17}, 135 (1983);
{\bf 21}, 291 (1984);
E.~van Beveren and G.~Rupp,
Eur.\ Phys.\ J.\ C {\bf 11}, 717 (1999)
[arXiv:hep-ph/9806248].

\bibitem{PRD21p772}
E.~van Beveren, C.~Dullemond, and G.~Rupp,
Phys.\ Rev.\ D {\bf 21}, 772 (1980)
[Erratum-ibid.\ D {\bf 22}, 787 (1980)].

\bibitem{PRD27p1527}
E.~van Beveren, G.~Rupp, T.~A.~Rijken, and C.~Dullemond,
Phys.\ Rev.\ D {\bf 27}, 1527 (1983).

\bibitem{PLB592p1}
S.~Eidelman {\it et al.} \/[Particle Data Group Collaboration],
Phys.\ Lett.\ B {\bf 592}, 1 (2004).

\bibitem{PRD69p112002}
K.~Abe {\it et al.}  [Belle Collaboration],
Phys.\ Rev.\ D {\bf 69}, 112002 (2004)
[arXiv:hep-ex/0307021].
\end{thebibliography}
\end{document}